\documentclass[aps, prl, twocolumn]{revtex4}
\usepackage{graphicx}
\usepackage{amsmath}
\usepackage{natbib}
\usepackage{tikz}
\usepackage[caption=false]{subfig}

\begin{document}

\title{Phonon-Assisted Ultrafast Charge Separation in a Realistic PCBM Aggregate}
\author{Samuel L Smith}
\email{sls56@cam.ac.uk}

\author{Alex W Chin}
\affiliation{Cavendish Laboratory, University of Cambridge}

\begin{abstract}
Organic solar cells must separate strongly bound electron-hole pairs into free charges. This is achieved at interfaces between electron donor and acceptor organic semiconductors. The most popular electron acceptor is the fullerene derivative PCBM. Electron-hole separation has been observed on femtosecond timescales, which is incompatible with conventional Marcus theories of organic transport. In this work we show that ultrafast charge transport in PCBM arises from its broad range of electronic eigenstates, provided by the presence of three closely spaced delocalised bands near the LUMO level. Vibrational fluctuations enable rapid transitions between these bands, which drives an electron transport of $\sim$3 nm within 100 fs. All this is demonstrated within a realistic tight binding Hamiltonian containing transfer integrals no larger than 8 meV.
\end{abstract}

\maketitle

The best solution processed organic photovoltaic cells now exhibit efficiencies exceeding 9$\%$ \cite{He2012a}. Photons are absorbed within these devices generating bound excitons. These excitons must be separated into free charges at interfaces between electron donor and acceptor semiconductors. However in order to separate, electron and hole must overcome their mutual Coulomb attraction, which is typically an order of magnitude greater than $k_BT$ at room temperature \cite{Gelinas2011}. Experimentally, charge separation has been observed on ultrafast timescales ($<$100 fs) in a range of devices \cite{Bakulin2012, Grancini2012, Jailaubekov2012}. This observation is incompatible with conventional theories of localised charge transport in organic media, and a number of new proposals have emerged \cite{Troisi2013, Caruso2012, Tamura2013, Bittner2014}.

\textit{G$\acute{e}$linas et al.} found that ultrafast charge separation was only observed in devices containing nanoscale aggregates of PC$_{71}$BM, the chosen electron acceptor \cite{Gelinas2014}. Alongside this experiment we presented a simple phenomenological model of ultrafast transport through the delocalised states of these acceptor aggregates. We concluded that the acceptor LUMO bandwidth must be at least similar in magnitude to the electrostatic binding energy of electron and hole across the interface ($\sim$0.3 eV). This condition was necessary to ensure that electronic states could delocalise across both interface and bulk; generating an effective coupling between delocalised bulk states and the localised charge separation event.

In another recent work, \textit{Savoie et al.} performed detailed DFT simulations on small ($\sim$5$^3$ nm$^3$) aggregates of PC$_{61}$BM \cite{Savoie2014, Cheung2010}. Unlike the effective model considered in ref. \cite{Gelinas2014}, the fullerene LUMO is composed of three degenerate states. In functionalised PCBM this degeneracy is split into three nearby levels, which give rise to three closely spaced electron bands near the LUMO level. While the width of any single band is $\sim$0.1 eV, these three bands in combination give rise to a broad range of electronic states near the interface. The combined width of these states is $\sim$0.4 eV, consistent with the phenomenological condition described above. The goal of this paper is to demonstrate how these three bands combine to drive ultrafast charge transport in a realistic model of a PCBM aggregate. We will find that ultrafast transport arises from rapid interband transitions between states at similar energies, and that these transitions are enabled by scattering within the surrounding vibrational environment. Thus our work provides another example of noise-assisted dynamics; which has recently been highlighted in biological light-harvesting \cite{Mohseni2008,Plenio2008,Lambert2013,Scholes2011}.

To model the PCBM eigenstates, we form the tight binding Hamiltonian,
\begin{eqnarray}
H_S = && \sum_{i,b} E_{i,b}|i,b><i,b| + J_{int} \sum_{b,i,j}(|i,b><j,b| + h.c.) \nonumber \\
    && + J_{ext} \sum_{b'\neq b, i,j}(|i,b><j,b'| + h.c.).   
\end{eqnarray}
The indices b and b' label the three electronic levels on each PCBM molecule, while the indices i and j label the location of a molecule within the aggregate. Summations over j are performed only over nearest neighbours of the site i. J$_{int}$ labels the transfer integral between nearest neighbours of the same band, while J$_{ext}$ labels the coupling between nearest neighbours of different bands. For simplicity we choose a primitive cubic unit cell, with lattice parameter 1 nm and aggregate size 7$^3$ nm$^3$. Based on the results of \textit{Savoie et al.}, we take J$_{int} = 8$ meV, and J$_{ext} = 3$ meV. 

We assume that immediately after charge separation the hole lies adjacent to the aggregate, and we incorporate the Coulomb potential felt by the electron into the site energies $E_{i,b} = \Delta_b + \sigma_{i,b} - e^2/(4\pi \epsilon_0 \epsilon_r r_i)$. $r_i$ labels the distance between the hole and the i$^{\text{th}}$ PCBM molecule. The dielectric constant $\epsilon_r = 4$, and the minimum separation between electron and hole $r_1$ = 1 nm. $\sigma_{i,b}$ labels a random uncorrelated Gaussian disorder, with standard deviation $\sigma = 10$ meV. Finally $\Delta_b$ describes the energy offsets of the 3 states near the LUMO level, which generates the splitting between the three delocalised bands. We choose $\Delta_1 = 0$ meV, $\Delta_2 = 80$ meV and $\Delta_3 = 300$ meV.

By diagonalising this Hamiltonian we may calculate the density of states within the aggregate, both including and neglecting the nearby hole. These are plotted in figures 1a-b. By comparison with figures S1a-b in ref \cite{Savoie2014} it can be seen that this model is compatible with more detailed treatments of small PCBM aggregates. However to explore how these eigenstates may lead to ultrafast charge transport, we must consider both their energies and their spatial structure.

In figure 1c we plot the state energies against their mean separation distance from the hole. This plot emphasizes the presence of three electronic bands, whose eigenstate energies are suppressed near the interface. Exciton dissociation will inject electrons onto molecules near the hole. While the corresponding eigenstates of the 1$^{\text{st}}$ and 2$^{\text{nd}}$ band are localised within the Coulomb well, the relevant states of the 3$^{\text{rd}}$ band lie at the same energy as more delocalised states in the aggregate bulk. Thus ultrafast charge separation can, in principle, be driven by electron transfer between localised interfacial states and isoenergetic delocalised bulk states \textit{arising from different bands}.

Since the bands are weakly coupled, we might ask whether purely coherent dynamics can enable charge separation. In figure 1d we plot the coherent propagation of a charge within the aggregate. At time zero the electron is localised on the third energy level of the PCBM molecule adjacent to the hole. Coherent propagation generates electron-hole separations of $\sim$2 nm, but on a picosecond timescale. In the absence of disorder this separation is suppressed, which indicates that the transport is driven by a gradual loss of phase coherence as the electron repeatedly scatters within the aggregate.

\begin{figure}[t]
\subfloat{\includegraphics[scale = 0.5]{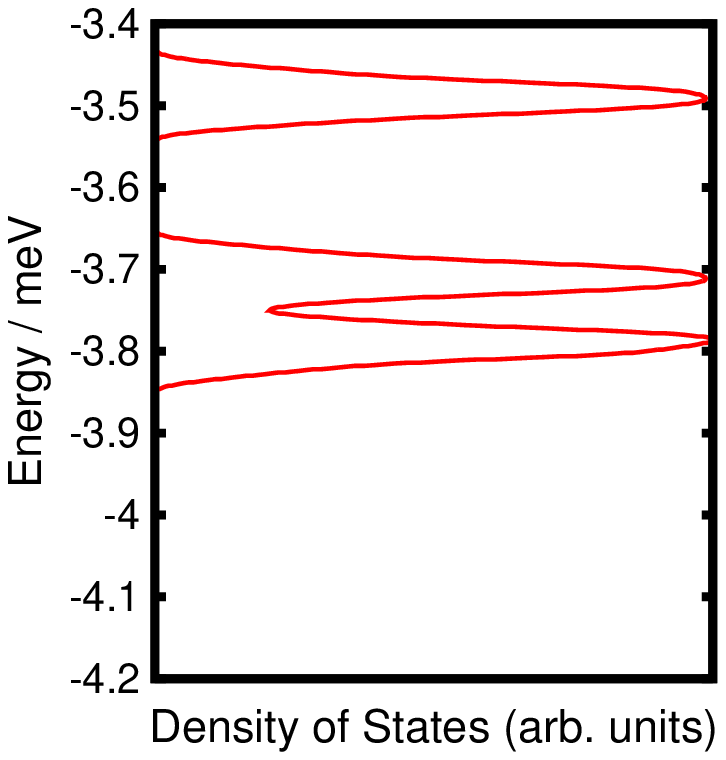}}
\subfloat{\includegraphics[scale = 0.5]{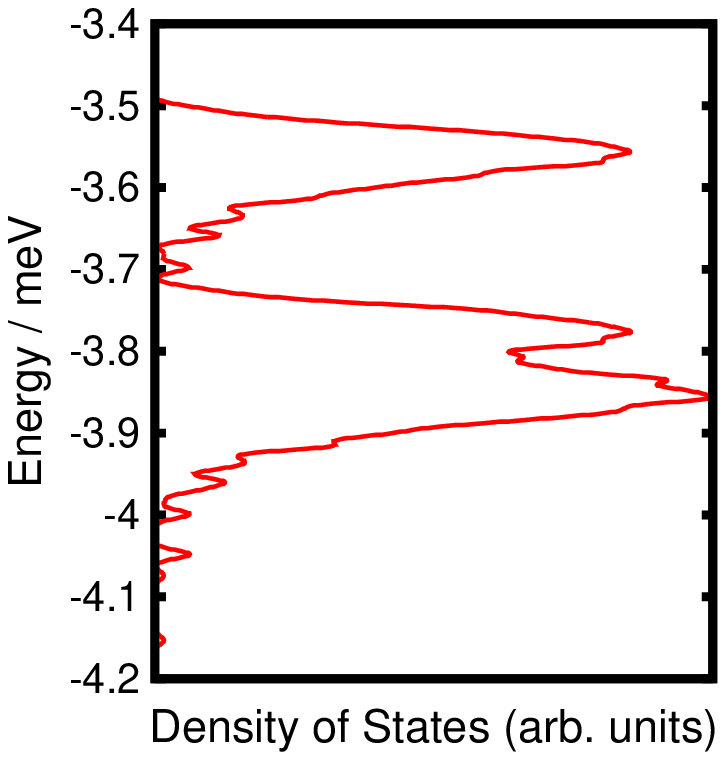}} \\
\subfloat{\includegraphics[scale = 0.5]{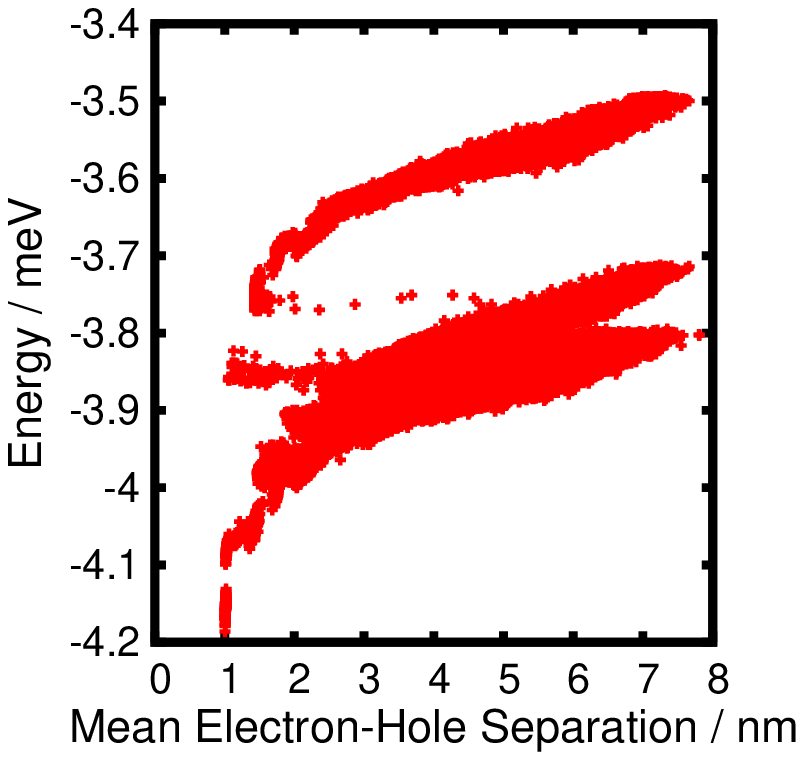}}
\subfloat{\includegraphics[scale = 0.5]{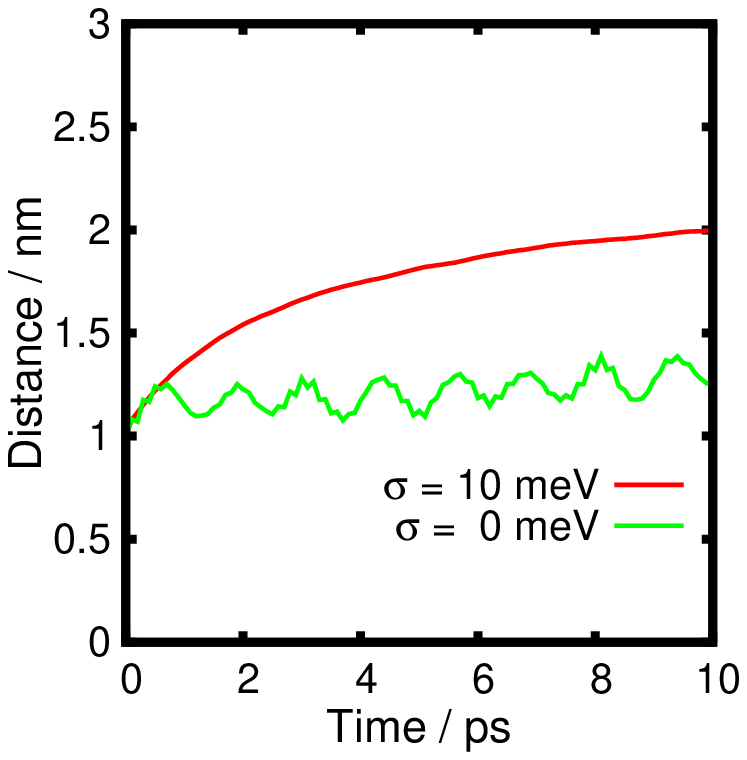}}
\caption{Top left (a): The density of states of a 7$^3$ nm$^3$ aggregate, neglecting the nearby hole. Top right (b): The density of states, when the hole is introduced 1 nm away from the centre of one face of the aggregate. We neglect disorder for best comparison with ref \cite{Savoie2014}, and each eigenstate is convolved with a 5 meV Gaussian to generate a continuous distribution. Bottom left (c): A plot of eigenstate energies and their electron-hole separations. Each point corresponds to a single eigenstate, and the eigenstates of 100 disorder realizations are shown. Eigenstate energies are suppressed near the hole. This brings the interfacial states of the 3$^{\text{rd}}$ band into resonance with delocalised bulk states of the lower two bands. Interband coupling enables mixing between the bands near the interface. Bottom right (d): Mean electron-hole separation for coherent propagation of an electron, which initially lies on the 3$^{\text{rd}}$ level adjacent to the hole (averaged over 1000 simulations). Interband coupling enables separation to 2 nm, but this occurs on a picosecond timescale.}
\end{figure}

In order to drive charge transport on femtosecond timescales, we must introduce a perturbation to our Hamiltonian which can drive transitions between localised states at the interface and the degenerate delocalised states in the bulk. Such a perturbation arises from the molecular vibrations, which we neglected in our initial Hamiltonian. We incorporate these vibrations within a linear-vibronic model, such that vibrations induce fluctuations in the site energies of each molecule. The Hamiltonian of the total vibronic system is given by \cite{Weiss1999, May2008, Holstein1959, Bittner2014},
\begin{eqnarray}
H & = & H_S + H_B + H_I, \\
H_B & = & \sum_{i,b,k} \omega_k a_{ibk}^\dagger a_{ibk}, \\
H_I & = & \sum_{i,b,k} (g_k a_{ibk}^\dagger + g_k^*a_{ibk}) |i,b><i,b|. 
\end{eqnarray}
For simplicity, each energy level on each site has been coupled to its own independent set of molecular vibrations. The molecular modes can be characterized by the spectral function $J(E) = \sum_k |g_k|^2 \delta(E-\hbar w_k)$, which we assume is identical for each molecule. The electronic dynamics in the presence of molecular modes can be treated within the secular Redfield theory \cite{redfield1957theory, May2008, Pollard2007}; which enables us to propagate the reduced electron density matrix. The eigenstate populations evolve according to a set of transition rates,
\begin{eqnarray}
\frac{d\rho_{ii}(t)}{dt} & = & \sum_{j\neq i} R_{ji}\rho_{jj}(t) - R_{ij}\rho_{ii}(t), \\
R_{ij} & = & 2\pi J(|E_i-E_j|) n(E_i-E_j) \sum_{a,b} |C_{ab}^i C_{ab}^j|^2.
\end{eqnarray}
The $C_{ab}^i$ coefficients label the site expansion of the i$^{\text{th}}$ electronic eigenstate $|i> = \sum_{a,b} C_{ab}^i|a,b>$, while the thermal function $n(E) = 1/(Exp(E/k_BT)-1)$ if $E<0$ and $n(E) = 1$ if $E>0$. Note that R$_{ij}$ represents Fermi's Golden Rule for single phonon transitions between electronic eigenstates, while Redfield theory assumes that the vibrational bath returns to thermal equilibrium before any subsequent transition occurs. Coherences in the density matrix decay according to
\begin{eqnarray}
\rho_{ij}(t) & = & \rho_{ij}(0)e^{-\gamma_{ij}t - i(E_i-E_j)t/\hbar}, \\
\gamma_{ij} & = & \sum_k (R_{ik} + R_{jk})/2.
\end{eqnarray}
The contributions $(R_{ii} + R_{jj})/2$ in the summation above introduce pure dephasing. One of the simplest spectral functions that we can consider is the Drude function of a single overdamped mode, $ J(E) = \Delta E \gamma / (E^2 + \hbar^2 \gamma^2)$ \cite{Mukamel1995}. $\Delta$ labels the reorganization energy, while $\tau = 2\pi/\gamma$ labels the damping timescale. Large $\tau$ represents a vibrational bath dominated by low frequency vibrations, while small $\tau$ represents a vibrational bath with a broad band of weakly coupled modes. The Drude spectral density is often applied to biological photosynthetic systems, for which damping timescales of $\tau \sim$ 100 fs are typical \cite{Ishizaki2009a}. Following \textit{Savoie et al.}, we estimate a reorganization energy of 15 meV. 

In figure 2a, we present the evolution of the electron-hole separation at 300 K. Considering first the curve $\tau =$ 100 fs, we see that the molecular vibrations have enabled the electron to reach 3.5 nm away from the hole within 100 fs. This represents the main result of this paper. Since we have not allowed the electron to escape from the aggregate, on longer timescales it is drawn back towards the interface as the eigenstate populations approach thermal equilibrium. The curve $\tau =$ 1000 fs suggests that ultrafast transport is primarily driven by the low frequency modes, which enable transitions between states similar in energy. As anticipated by our earlier discussion, these enable long-range tunneling between interfacial and delocalised states. The high frequency modes allow the electron to collapse rapidly into trapped states at the interface, thus preventing ultrafast charge separation and enhancing the rate of thermalisation. Within this simple treatment of the vibrational bath, the fate of injected electrons is thus fairly sensitive to the shape of the spectral function.

Figure 2b presents the electronic dynamics at 4 K. The ultrafast transport is suppressed, since transitions now primarily occur to lower energy states. However charge separations $>$ 2.2 nm are still reached within 100 fs. Once again, transport is enhanced in the presence of low frequency modes. As we discuss below, it is likely that the transport at low temperatures would be further enhanced if the assumptions underlying the derivation of Redfield theory were relaxed. We note that ultrafast charge transport was observed at 4 K by \textit{G$\acute{e}$linas et al.} \cite{Gelinas2014}.

We note here that while ultrafast charge transport has been observed in many devices, it may or may not be crucial to device performance. Ultrafast transport will generate free charges if the timescale for a delocalised charge to escape from an aggregate is shorter than the time required for the charge to thermalise back into the trapped interfacial states \cite{Groves2008}. Within the simulations above, this timescale is a few picoseconds \footnote{Note that we have neglected the possibility of polaron formation here, in order to focus on the ultrafast dynamics. Incorporating polaron formation may affect the thermalisation timescale.}. If a delocalised charge does relax back into the trapped states, then either electron and hole will recombine, or they may be separated at later times by thermal fluctuations \cite{Vandewal2014a, Groves2013a}. 

Perhaps the most important assumption underlying Redfield theory states that the vibrational modes remain in thermal equilibrium at all times. This assumption is reasonable if the timescale for the vibrations to regain equilibrium, given approximately by $\tau$, is shorter than the typical transition time between eigenstates. Although frequently applied, this assumption is rather questionable in this context. \textit{Tamura and Burghart} have shown that relaxing this approximation can enhance ultrafast transport \cite{Tamura2013}; electronic transitions to lower energy states excite vibrational modes, these excited vibrational modes can then promote the electron to other higher lying states away from the interface. Such effects would be particularly beneficial to ultrafast transport at low temperatures.

\begin{figure}[t]
\subfloat{\includegraphics[scale = 0.5]{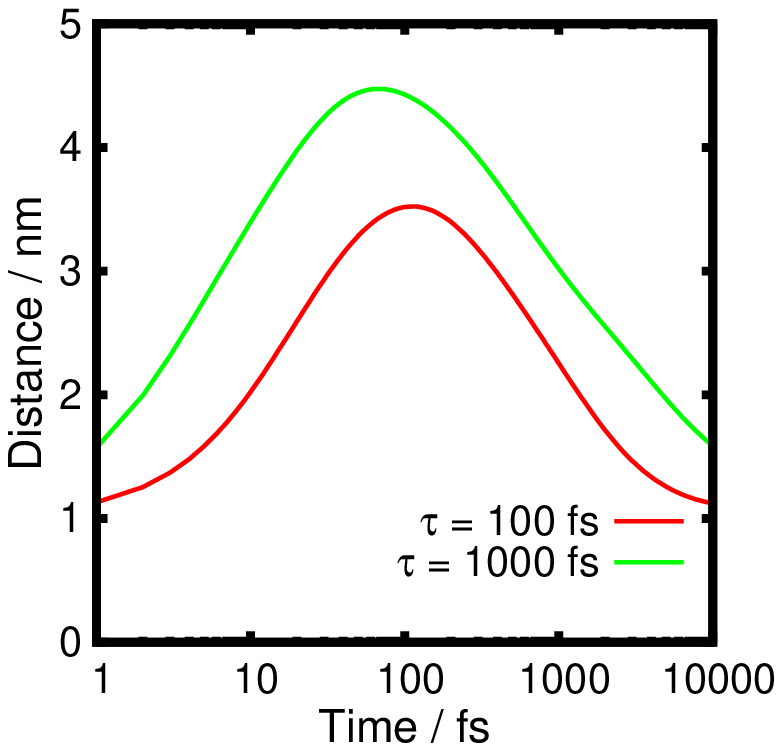}}
\subfloat{\includegraphics[scale = 0.5]{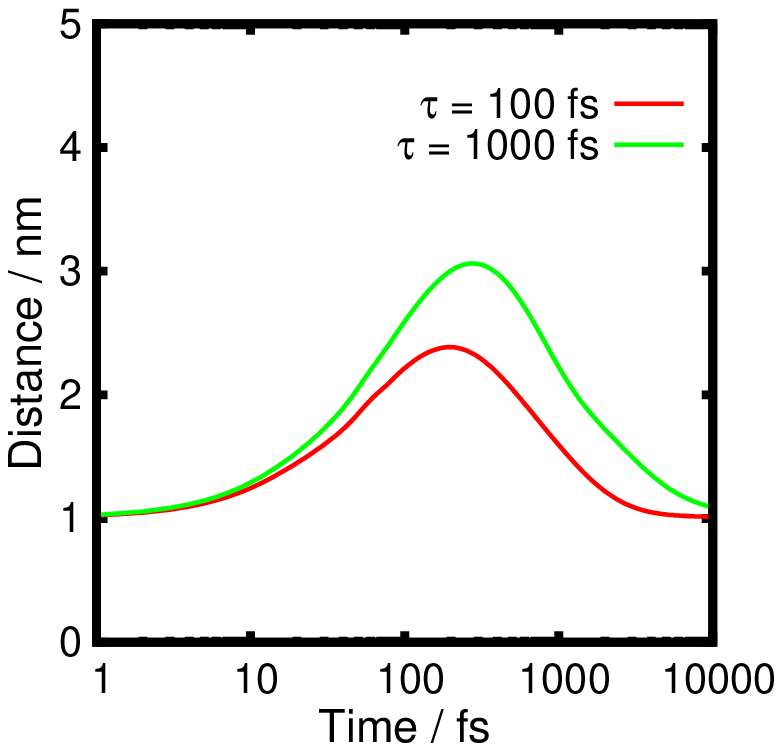}}
\caption{Top left (a): Mean electron-hole separation under Redfield dynamics at 300 K, for damping timescales $\tau =$ 100 and 1000 fs. Note the log scale. At early times the electron propagates rapidly into the aggregate. For both cases considered the electron-hole separation reaches 3.5 nm within 100 fs. Ultrafast charge separation is enhanced when $\tau =$ 1000 fs, since the spectral density is dominated by low frequency modes linking the injection site to delocalised states in the bulk. Since the electron cannot escape the aggregate, at long times the eigenstate populations approach thermal equilibrium. As a result, the electron collapses back towards the interface on a ps time scale. Top right (b): Mean electron-hole separation under Redfield dynamics at 4 K. Ultrafast charge separation is suppressed, and rate of thermalisation back towards the interface is enhanced. Never-the-less, electron-hole separations of 2.2 nm are still reached within 100 fs. All simulations are averaged over 1000 disorder realizations.}
\end{figure}

The two remaining assumptions are the Markov and Secular approximations, applied within 2$^{\text{nd}}$ order perturbation theory. Detailed discussion of these approximations can be found elsewhere \cite{Ishizaki2009, Egorova2003}, here we note that they are in essence the approximations behind Fermi's Golden Rule. They imply that the transition rate $R_{ij}$ between two eigenstates, which is in principle a time dependent function of the coherences between eigenstates, can be replaced by a constant value. Such a replacement also implies that transitions can only occur which precisely conserve energy, in contradiction to the energy-time uncertainty relation. Relaxing such approximations would enable zero phonon fluctuations between localised interfacial states and nearly degenerate bulk states. Once again, we anticipate that Redfield theory should provide a pessimistic account of the ultrafast dynamics, thus justifying its use here. However, we note that Redfield dynamics averages over fluctuations in the vibrational bath. Strictly speaking we should not consider the dynamics on timescales shorter than typical vibrational timescales ($\sim\tau$). Consequently the ultrafast dynamics when $\tau = 1000$ fs is a little suspect; but is included to illustrate the beneficial impact of low frequency modes. Methods which relax the approximations above exist, but present a major numerical challenge in systems of this size \cite{Egorova2003, Wang2003, Mccutcheon2011, Makri1995, Prior2010, Chin2013, Ishizaki2009a}. In spite of the caveats above, our qualitative picture of ultrafast transport, driven by interband scattering between nearly degenerate states, should be robust.

Our mechanism for ultrafast charge transport depends crucially on the presence of nanoscale PCBM aggregates; in figure 3a we present the electron-hole separation for a range of aggregate sizes, taking $\tau =$ 100 fs and $T =$ 300 K. The larger the aggregate, the further apart the electron and hole can separate and the slower the process of thermalisation back towards the interface. In real devices, the optimum aggregate size will be determined by the competition between a number of conflicting factors, such as the exciton diffusion length and the need to preserve percolation pathways through the device. In figure 3b, we explore the dependence of ultrafast charge transport on the band splitting $\Delta_3$. This splitting is determined by the choice of side chain used to functionalise fullerene. The optimal splitting lies between 260-300 meV; which matches the band splitting observed in PCBM, and may help explain the superiority of PCBM to other fullerene derivatives. Further optimization might be achieved by varying the intermediate band offset $\Delta_2$.

\begin{figure}[t]
\subfloat{\includegraphics[scale = 0.5]{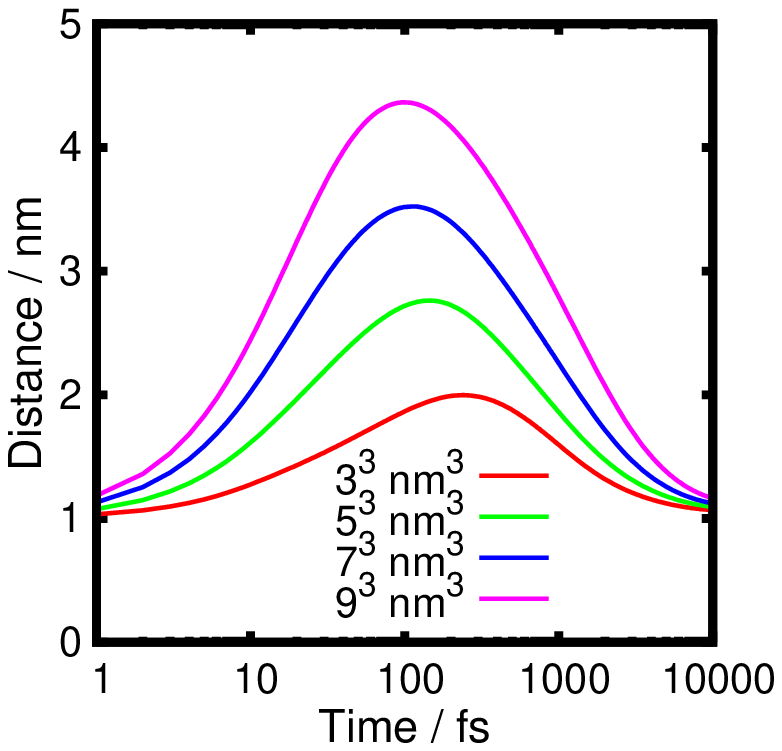}}
\subfloat{\includegraphics[scale = 0.5]{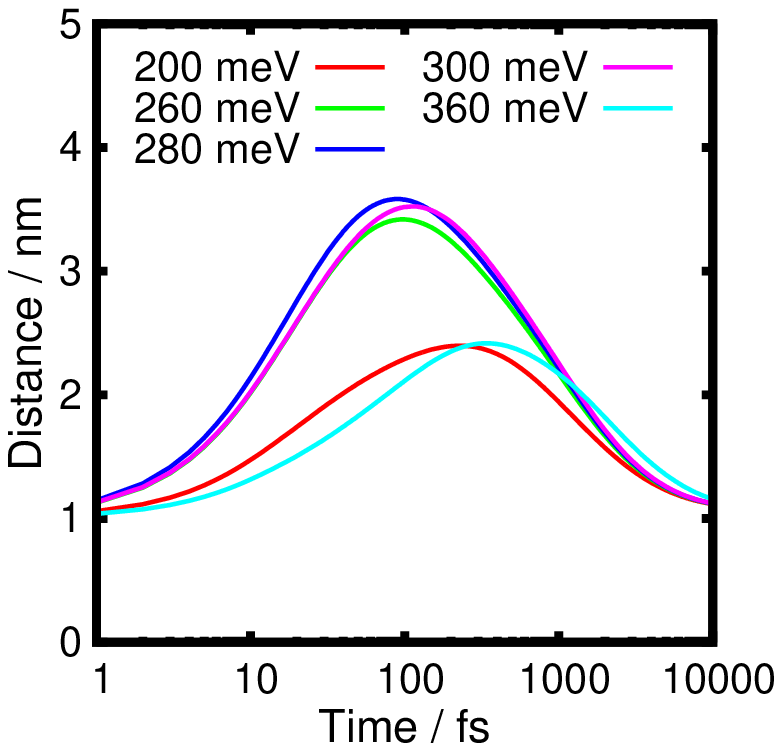}} \\
\subfloat{\includegraphics[scale = 0.5]{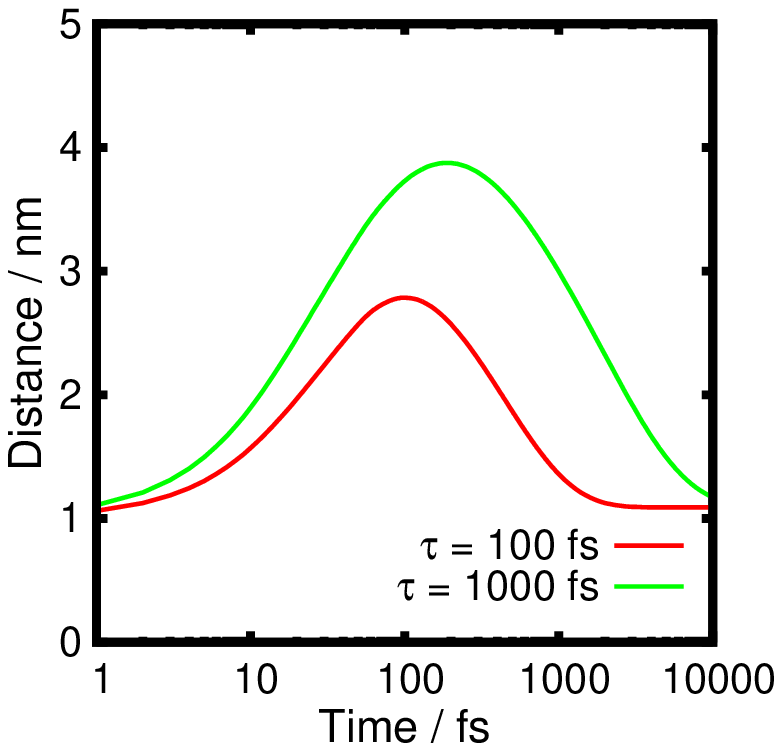}}
\subfloat{\includegraphics[scale = 0.5]{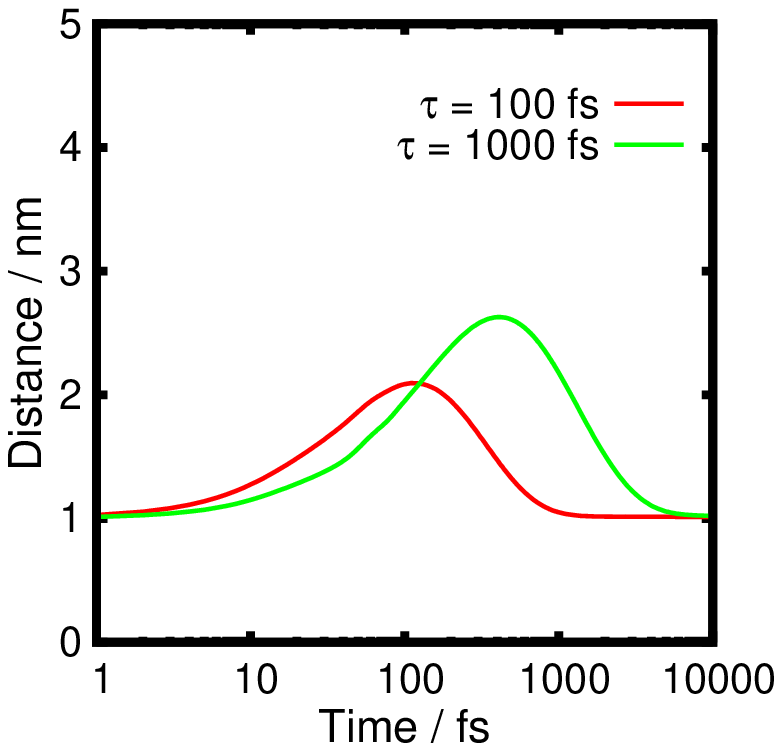}}
\caption{Top left (a): Electron dynamics at 300 K, with $\tau = 100$ fs, for a range of aggregate sizes. The larger the aggregate the further apart electron and hole separate at early times. Top right (b): Electron dynamics for a range of band offsets $\Delta_3$. The largest separations at 100 fs are achieved for offsets in the range $\Delta_3 =$ 260-300 meV. Bottom left (c): electron-hole separation at 300K under fluctuations in the off-diagonal couplings between states. Bottom right (d): electron-hole separation at 4 K under off-diagonal fluctuations. All simulations are averaged over 1000 disorder realizations.}
\end{figure}

Alongside the molecular vibrations treated above, organic crystals also exhibit lattice vibrations. These lattice vibrations induce large fluctuations in the transfer integrals between nearest neighbours. Troisi and Orlandi showed that these fluctuations can be similar in scale to the mean value \cite{Troisi2006}. Lattice vibrations might be expected to enhance the ultrafast dynamics, especially since many such modes lie at low frequency; however they are difficult to include here since they should not be treated as a perturbation \cite{Cheung2010}. Semi-classical treatments of these vibrations are often problematic, since these do not reach thermal equilibrium at long times \cite{Ciuchi2011}.

A full treatment of this problem is beyond the scope of this work. However to illustrate their potential importance, we briefly explore the effect of lattice fluctuations within the Redfield theory. We neglect the molecular vibrations, and couple the transfer integrals to a set of three vibrational modes on each lattice site. The interaction Hamiltonian and the Redfield transition rates are given in the appendix. Once again we use the Drude spectral density, and we take the reorganization energy to be one tenth of the mean transfer integral \footnote{The reorganization energy is different for fluctuations of intraband and interband transfer integrals.}. The electron dynamics at both 300K and 4K is plotted in figures 3c-d. We see that lattice fluctuations are also able to drive ultrafast charge separation, though we emphasize that a proper treatment of these modes should not be perturbative. We speculate that these lattice vibrations could possibly dominate over the molecular vibrations in real systems.

In conclusion, ultrafast charge transport in PCBM arises from the presence of three closely spaced conduction bands near the LUMO level. The Coulomb well provided by the nearby hole mixes these three bands, generating an effective bandwidth of $\sim$0.4 eV near the interface. Molecular and lattice vibrations are coupled to the electronic states, driving rapid transitions between localised interfacial states and delocalised states in the bulk. We have shown that a simple but realistic model of small PCBM aggregates, containing transfer integrals no larger than 8 meV, can generate electron hole separations $>$ 3 nm within 100 fs.

\appendix*
\section{Appendix}

Fluctuations in the transfer integrals between neighboring sites are introduced through the interaction Hamiltonian $H_I = H_I^x + H_I^y + H_I^z$, where
\begin{eqnarray}
&& H_I^z = \sum_{l,m,n,k,b} \left( g_k a^{\dagger}_{lmnk} + g^*_k a_{lmnk} \right) \times \nonumber \\
&& \left( |l,m,n,b \rangle \langle l,m,n+1,b| - |l,m,n-1,b \rangle \langle l,m,n,b| + h.c. \right) +  \nonumber \\
&& \sqrt{\frac{J_{ext}}{J_{int}}} \sum_{l,m,n,k,b' \neq b} \left( g_k a^{\dagger}_{lmnk} + g^*_k a_{lmnk} \right) \times \nonumber \\ 
&& \left(|l,m,n,b \rangle \langle l,m,n+1,b'| - |l,m,n-1,b' \rangle \langle l,m,n,b| + h.c.   \right) . \nonumber
\end{eqnarray}
l, m and n label the location of a molecule in three dimensions. In the main text these were grouped into a single label i. $H_I^x$ and $H_I^y$ can be readily obtained from $H_I^z$ by symmetry. The three components of $H_I$ are each accompanied by their own independent set of vibrational modes $H_B$ (defined in the main text). 

The transfer rates arising from a single component of the interaction Hamiltonian $H_I^z$ are given by
\begin{eqnarray}
R_{ij}^z = 2\pi J(|E_i-E_j|) n(E_i-E_j) \sum_{l,m,n} |M_{lmn}^{ij}|^2, \nonumber
\end{eqnarray}
where
\begin{eqnarray}
&& M_{lmn}^{ij} = 2 \sum_b \left( C_{l,m,n,b}^i C_{l,m,n+1,b}^j - C_{l,m,n-1,b}^i C_{l,m,n,b}^j \right) + \nonumber \\
&& 2 \sqrt{\frac{J_{ext}}{J_{int}}} \sum_{b' \neq b} \left( C_{l,m,n,b}^i C_{l,m,n+1,b'}^j - C_{l,m,n-1,b'}^i C_{l,m,n,b}^j \right). \nonumber
\end{eqnarray}
The factor of two accounts for hermitian conjugates, since all the coefficients $C_{l,m,n,b}^i$ are real. Note that terms in $M_{lmn}^{ij}$ arising from coefficients $C_{l,m,n,b}^i$ lying outside of the aggregate are set to zero (hard boundary conditions). The full rate $R_{ij} = R_{ij}^x + R_{ij}^y + R_{ij}^z$. Once again, $R_{ij}^x$ and $R_{ij}^y$ can easily be found by symmetry.

As for the molecular vibrations, we use the Drude spectral density $J(E) = \Delta E \gamma / (E^2 + \hbar^2 \gamma^2)$. We take the reorganization energy $\Delta$ to be 0.8 meV, one tenth of the intraband transfer integral. The scaling factor $\sqrt{\frac{J_{ext}}{J_{int}}}$ in the equations above ensures that this is reduced to 0.3 meV for fluctuations in the interband couplings.

\bibliography{library2}

\end{document}